# The Electronic Structure of Few-Layer Graphene: Probing the Evolution from a 2-Dimensional to a 3-Dimensional Material

Kin Fai Mak[1], Matthew Y. Sfeir[2], James A. Misewich[2], and Tony F. Heinz[1*]

[1]*Departments of Physics and Electrical Engineering, Columbia University, New York, NY 10027, USA*
[2] *Brookhaven National Laboratory, Upton, NY 11973, USA*

One Sentence Summary

The development of the electronic structure from 2-dimensional mono- and few-layer graphene to 3-dimensional bulk graphite was examined systematically by infrared absorption spectroscopy and successfully described by a unified zone-folding scheme applied to the graphite band structure.

Abstract

While preserving many of the unusual features of single-layer graphene, few-layer graphene (FLG) provides a richness and flexibility of electronic structure that render this set of materials of great interest for both fundamental studies and applications. Essential for progress, however, is an understanding of the evolution of the electronic structure of these materials with increasing layer number. In this report, the evolution of the electronic structure of FLG, for $N = 1, 2, 3, \ldots, 8$ atomic layers, has been characterized by measurements of the infrared conductivity spectra $\sigma(\hbar\omega)$. For each layer thickness $N$, distinctive peaks are found in $\sigma(\hbar\omega)$, with positions obeying a simple scaling relation. The observations are explained by a unified zone-folding scheme that generates the electronic structure for all FLG materials from that of bulk graphite.

*Corresponding author: tony.heinz@columbia.edu

Graphene, a single-monolayer of $sp^2$-hybridized carbon, has attracted much attention because of its unique electronic properties (*1*). In view of the potential for band-structure engineering (*2-5*) and new applications (*6*), recent interest has also focused on graphene's multilayer counterparts. In this respect, the most basic issue is to develop an understanding of the electronic structures of few-layer graphene (FLG) as it evolves from the characteristic linearly dispersing two-dimensional bands of single-layer graphene towards the band structure of bulk graphite (*7*). Recent theoretical studies (*8-16*) have predicted dramatic changes in the low-energy electronic properties: When two or more layers of graphene are present, the linearly dispersing bands are either replaced or augmented by split hyperbolic bands. The overall electronic structure changes sensitively with increasing number of layers *N*, ultimately approaching the bulk limit of graphite. Despite these fascinating predictions, experimental investigations have been limited to single-layer graphene and a few recent investigations of the electronic properties of bilayer graphene (*17-20*).

In this paper, we trace experimentally the evolution of the electronic structure of FLG from the ideal 2-dimensional (2-D) sheet of single-layer graphene towards the 3-dimensional (3-D) bulk solid. Measurements of the low-energy optical conductivity $\sigma(\hbar\omega)$ over the photon energy range 0.2 – 0.9 eV reveal a distinct and well-structured spectrum for each layer thickness of $N$ = 1, 2, 3, … 8 layers. The pronounced resonant features observed in $\sigma(\hbar\omega)$ reflect the critical points in the band structure for each layer thickness and are found to obey a simple scaling behavior with *N*. We show that these observations can be understood within a tight-binding (TB) model of the electronic structure and that the *2-D band structure of all FLG of any layer thickness N* can be derived from the *3-D band structure of graphite* using a precisely defined zone-folding scheme to incorporate quantum confinement perpendicular to the planes. This simple, but rigorous analysis provides valuable insight into how the electronic structure and optical transitions evolve with increasing layer thickness from single-layer graphene to bulk graphite. The results also provide a spectroscopic basis for the identification of ordered graphene multilayer samples of differing thickness.

For direct interpretation of our optical data, we investigated graphene samples supported on bulk transparent substrates ($SiO_2$ from Chemglass, Inc). The FLG samples were deposited by mechanical exfoliation of kish graphite (Toshiba) on the substrates that had been carefully cleaning by sonication in methanol. The typical area of our graphene samples varied from several hundreds to thousands of $\mu m^2$. We established the thickness of each of the deposited graphene samples using absorption spectroscopy in the visible spectral range. This method provides a precise identification of layer thickness *N* for 10 or more monolayers, since each graphene monolayer absorbs approximately 2.3% (*21, 22*) of the light in this spectral range.

For each FLG sample of thickness *N* = 1, 2, 3, … 8, we obtained the optical conductivity spectra $\sigma(\hbar\omega)$ in the mid- to near-infrared. This spectral range covers the distinctive features associated with interband transitions. The measurements were performed using the National Synchrotron Light Source at Brookhaven National Laboratory (U2B beamline) as a bright source of broadband infrared radiation. We detected the reflected optical radiation with a Fourier-transform infrared (FTIR) spectrometer equipped with an HgCdTe detector under nitrogen purge. The measurements were performed in a reflection geometry. The synchrotron radiation could be focused to a spot size below 10 μm with a 32× reflective objective. The reflectance spectra of the graphene mono- and multilayers were obtained by normalizing the sample spectrum by that from the bare substrate, as in an earlier investigation of single-layer graphene (*22*).

For a sufficiently thin sample on a transparent substrate, $\sigma(\hbar\omega)$ of sample is related to the reflectance in a direct manner (*22*): The fractional change in reflectance associated with the presence of the thin-film sample is proportional to the real part of its optical sheet conductivity $\sigma(\hbar\omega)$, or equivalently, to its absorbance $A = (4\pi/c)\sigma(\hbar\omega)$. We can therefore convert the measured reflectance spectra into $\sigma(\hbar\omega)$ by multiplication of a suitable numerical factor determined by the (frequency-dependent) refractive index of the substrate (*22*). With increasing film thickness, propagation effects begin to play a role, and the relation between the dielectric function of the film and the measured properties becomes more complex. To determine the validity of the thin-film approximation in analysis of our data, we performed full calculations of the reflectance from films of finite thickness on a transparent substrate. Even for the thickest film considered in our measurements, the 8-layer graphene sample, we do not see significant changes in the shape of the inferred $\sigma(\hbar\omega)$; errors in the inferred magnitude of $\sigma(\hbar\omega)$ are limited to 15%. In the interest of simplicity, we consequently present all data based on the thin-film analysis.

$\sigma(\hbar\omega)$ of graphene samples for layer thicknesses of $N = 1, 2, \ldots, 8$ were obtained over a range of photon energies $\hbar\omega$ between 0.2 and 0.9 eV [Fig. 1]. The evolution of the FLG samples towards the bulk behavior is easier to appreciate in spectra that are normalized by the layer thickness $N$ [Fig. 2, left panel]. Before entering into a more detailed discussion of the features of the multilayer spectra and their relation to theoretical predictions, we would like to make several general observations about the experimental results. (1) In the limit of a single-layer thickness, we find significant absorption, but without any spectral features. Such universal absorption, with an absorbance of $\pi\alpha$ (where $\alpha$ denotes the fine-structure constant) or a conductivity of $\pi e^2/2h$, has been predicted theoretically (*23*) and has been recently demonstrated experimentally (*21, 22*). It arises from interband transitions between the linearly dispersing graphene bands. (2) For all graphene samples of thicknesses $N > 1$, we observe well-defined peaks in $\sigma(\hbar\omega)$. As we discuss below, these arise from van Hove singularities in the 2-D band structure of the multilayer system. (3) With increasing $N$, additional resonances appear in $\sigma(\hbar\omega)$, with characteristic energies evolving in a systematic manner (*24*). The trend can be visualized in a contour plot of the normalized sheet conductivity as a function of $N$ [Fig. 3]. The positions of the peaks are seen to follow a well-defined scaling behavior with layer thickness $N$. (4) The absolute strength of the resonances does not decrease significantly with layer number [Fig. 1]. On the other hand, $\sigma(\hbar\omega)$ normalized per layer [Fig. 2, left panel] show a clear evolution toward the behavior of bulk graphite, where a single broad feature present at approximately twice the energy of that seen in the bilayer sample.

To understand the multilayer graphene spectra $\sigma(\hbar\omega)$, we work within a TB description of their electronic structure. This type of model was introduced many years ago to describe the valence and conduction ($\pi$) bands of bulk graphite (*7*). In various simplified forms, it has been the basis of widely adopted descriptions of the $\pi$ bands of both single- and bilayer graphene (*8, 9*). Here we present a generalization that readily generates the band structure of single- and bilayer graphene, but also extends to multilayer graphene samples of arbitrary layer thickness. Our approach relies on a precisely defined zone folding of the 3-D graphite bands.

Let us consider a TB Hamiltonian for *N*-layer graphene with arbitrary couplings between atoms in the same or adjacent layers of graphene. We treat the electronic and structural properties of each layer in the sample (including the outer layers) as identical. This latter assumption is reasonable for graphene, given the saturated character of its in-plane bonds and the weakness of its out-of-plane interactions. In contrast, for most other material systems, such an approximation would introduce significant errors when applied in the limit of samples of just a few monolayers thickness where significant reconstruction would be expected.

Under the given hypotheses, the eigenstates of the Hamiltonian of *N*-layer graphene are simply a subset of those obtained for bulk graphite. The additional criterion that applies to the *N*-layer sample is that the wavefunctions must vanish at the position where the graphene planes lying immediately beyond the physical material, *i.e.*, at the positions of the $0^{th}$ and $(N+1)^{th}$ layer. This is accomplished by forming standing waves with the momentum perpendicular to the graphene planes quantized as

$$k_z = \frac{2\pi n}{(N+1)c}. \tag{1}$$

Here $c/2$ is the inter-layer separation (0.34 nm) and the allowed values for the index are $n = \pm 1, \pm 2, \pm 3, \ldots \pm(N+1)/2$ (for odd *N*) and $\pm N/2$ (for even *N*). Independent (standing-wave) states are generated only for positive values of *n* for which $k_z \leq \pi/c$, corresponding to the positive half of the graphite Brillouin zone. For *N*-layer graphene, the number of independent 2-D planes in the 3-D Brillouin zone, *i.e.*, the number of new 2-D bands being created, is thus given by the integer part of $(N+1)/2$.

To describe the 3-D $\pi$-bands of graphite, from which we then obtain the 2-D bands for FLG of any thickness, we make use of a simplified model that includes the nearest in-plane coupling coefficient, $\gamma_0$, and couplings between atoms in adjacent planes described by strengths $\gamma_1$, $\gamma_3$, and $\gamma_4$ (*7*). This Hamiltonian generates [Fig. 4] four bands along the H – K (*z*) direction: two degenerate bands ($E_3$) without any dispersion (since we have neglected coupling $\gamma_2$ as small compared to $\gamma_1$) and two dispersive bands ($E_1$ and $E_2$) with energies of $\pm 2\gamma_1 \cos(k_z c/2)$. The bands all become degenerate at the H point (since we have neglected coupling $\gamma_6$ as small compared to $\gamma_1$). The dispersion of the 2-D bands for FLG corresponds to that observed in planes perpendicular to the H – K direction at locations defined by the quantization condition (Eqn. 1) for $k_z$. At the H-point, we obtain (doubly-degenerate) bands that disperse linearly. For all other planes, the dispersion is described by split pairs of hyperbolic bands [Fig. 4]: two conduction bands dispersing upwards and two valence bands dispersing downwards.

For the graphene monolayer, the quantization criterion gives rise to only one allowed value for $k_z$, namely, $k_z = \pi/c$, corresponding to the H-point of the Brillouin zone. The zone-folding construction then yields immediately the well-known linearly dispersing bands of graphene. For the bilayer, the quantization criterion again yields just one independent value of the perpendicular momentum, $k_z = 2/3(\pi/c)$. The separation between the resulting hyperbolic bands is given by $2\gamma_1 \cos(\pi/3) = \gamma_1$. For FLG with $N > 2$, we obtain states like those of the bilayer, but with a modified splitting. In addition, for odd values of $N$ the construction includes the H-point, and we also generate the linearly dispersing bands of the monolayer. Thus for even $N$, we obtain $N/2$ sets of hyperbolic bands; for $N$ odd, we have a single linearly dispersing band and $(N-1)/2$ sets of hyperbolic bands. This solution to the electronic structure of FLG was presented earlier by Koshino and Ando in a slightly different formulation (*13*). From the dispersion relation for the graphite bands and quantization conditions, we see immediately the energy spacing of the hyperbolic bands in $N$-layer graphene hyperbolic bands is given (for positive integers $n \leq N/2$) by

$$\gamma_{N,n} = 2\gamma_1 \cos\left(\frac{n\pi}{(N+1)}\right). \qquad (2)$$

These energy differences are of particular interest to us. Given the peak in the joint density of states associated with the onset of the interband transition, we expect to observe a feature in $\sigma(\hbar\omega)$ at just these energies (*25, 26*).

With the band structure of FLG constructed as above, we can readily calculate the material's in-plane optical sheet conductivity, since only transition matrix elements between states with the same value of $k_z$ can contribute to the response (*27*). Thus, we need only consider transitions between the same sub-bands and can sum up the contributions of the effective bilayer components and, for FLG with $N$, an additional contribution from the set of linearly dispersing bands. The results of such a calculation [Fig. 2, right panel], carried out with the Kubo formula, reproduce the locations of the peaks observed experimentally for samples of layer thicknesses $N = 1, 2, 3, …, 8$. Some differences are apparent with respect to line shapes. These differences can be attributed largely to doping effects (supporting online text).

Within this theoretical framework, the experimentally observed scaling relationship for the different optical resonances [Fig. 3] is a direct consequence of transitions in the bilayer components of the each multilayer structure, as defined by Eqn. (2). The best fit to the experimental transition energies [Fig. 3] implies a value of $\gamma_1 = 0.37$ eV for the dominant interlayer coupling. This result is consistent with values reported in the literature (*7*). A few comments may also be made with respect to convergence to the bulk graphite limit. Carrying out the multilayer calculation in the limit of large $N$ does, in fact, yield a spectrum [Fig. 2, right panel] compatible with the experimental data for graphite. In the limit of large $N$, transitions from many different sub-bands must be considered within the given zone-folding schema. The peaks between the individual hyperbolic bands become closely spaced. However, unlike for the graphene monolayer, $\sigma(\hbar\omega)$ of graphite is not flat. Although transitions with energies less than $2\gamma_1$ are largely washed out in the limit of large *N*, the fact that the transitions energies saturate at $2\gamma_1$ leads to an increased conductivity around this energy (*28*). In terms of the bulk solid, this peak corresponds to the van Hove singularity at $2\gamma_1$ produced along the H – K direction of the 3-D graphite Brillouin zone.

In conclusion, the present work demonstrates that it is possible to develop a systematic understanding of the key features of the 2-D bands in *N*-layer graphene on the basis of a simple and precisely defined zone folding of the 3-D graphite bands. This situation is analogous to the canonical description of the 1-D bands of carbon nanotubes. The basic electronic structure of the panoply of different single-walled carbon nanotubes can be generated systematically from the 2-D electronic structure of graphene by zone folding (*29*). Similarly, the various types of electronic states in multilayer graphene are generated from the 3-D bands of graphite by an appropriate quantization condition.

**Figures:**

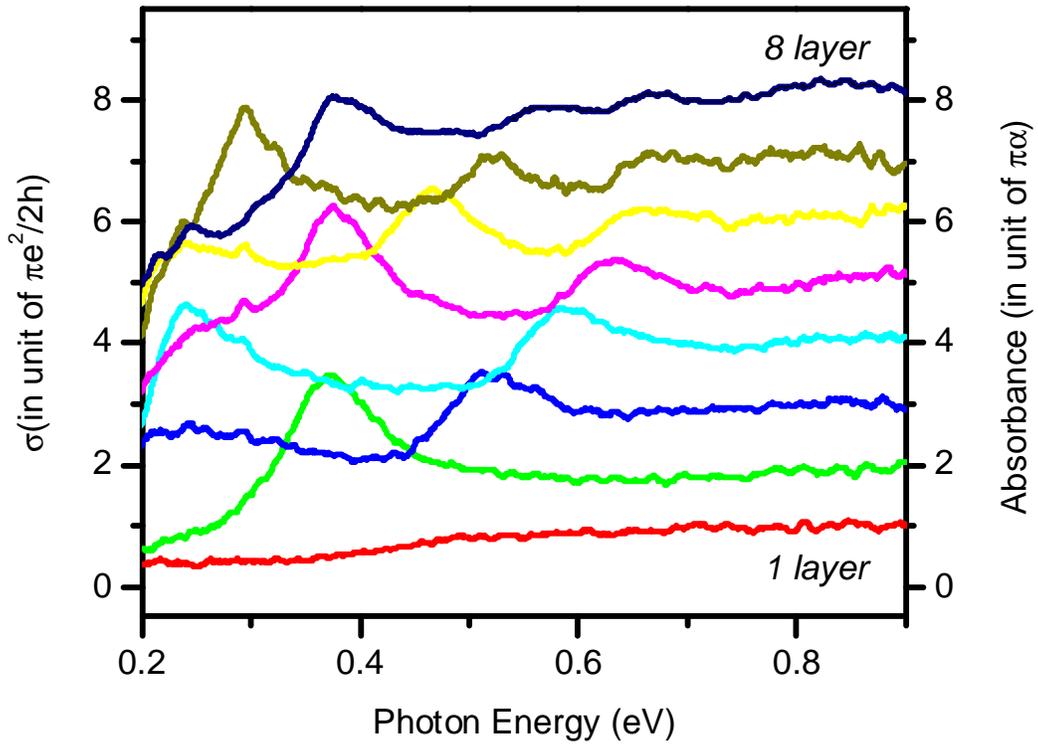

Fig. 1: Infrared (0.2 – 0.9 eV) conductivity/absorption spectra $\sigma(\hbar\omega)$ of multi-layer graphene samples with thickness from $N$ = 1 to 8 layers. The sheet conductivity and the absorbance are given in units of $\pi e^2/2h$ and $\pi\alpha$, respectively.

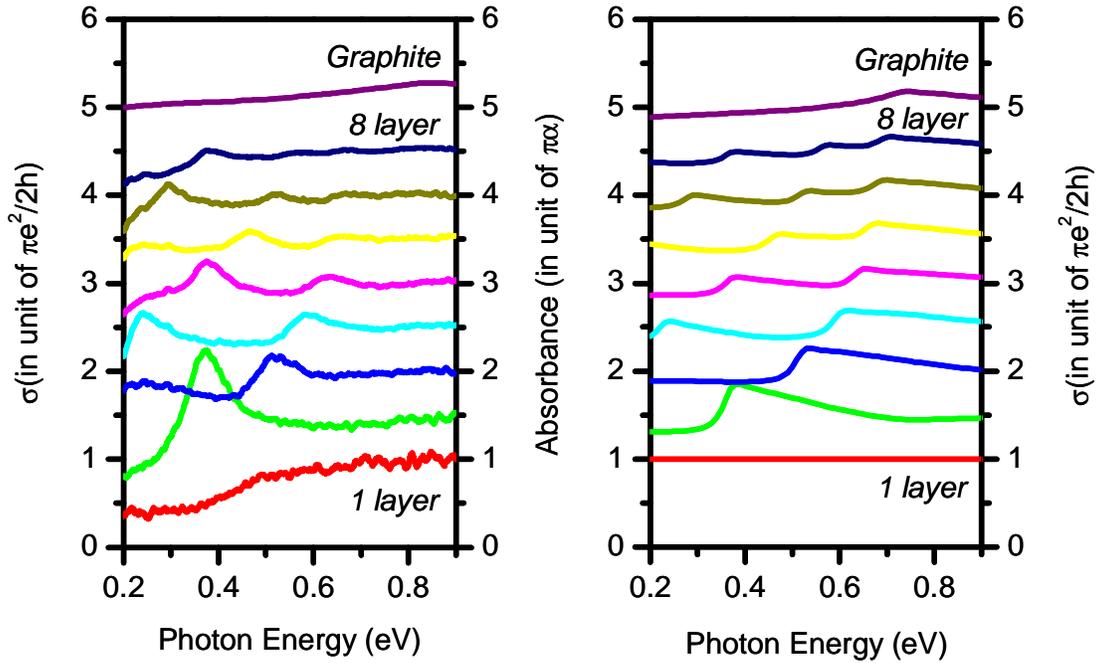

Fig. 2: The data of Fig. 1 for the sheet conductivity of *N*-layer graphene normalized to the layer thickness *N* and displaced for clarity of presentation. The top spectrum is the experimental result for graphite obtained from (*28*). We see the general convergence towards the graphite response with increasing thickness, although distinctive peaks are still observed to up *N* = 8. The right panel shows the calculated conductivity spectra using the Kubo formula and the zone-folding construction for the electronic states as described in the text.

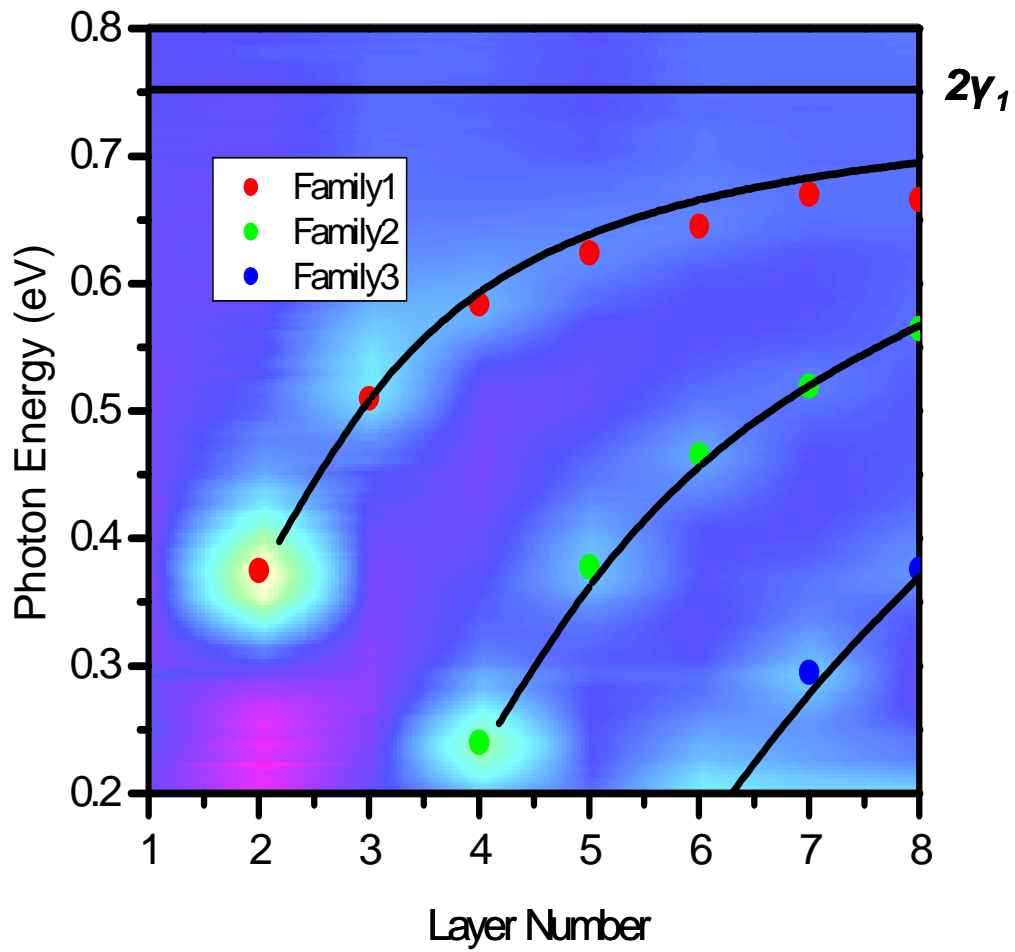

Fig. 3: Contour plot of the sheet conductivity per layer as a function of photon energy and number of layers *N*. Dots locate peak positions of the optical resonances in the experimental data. The positions can be grouped into three families that scale smoothly with layer number. The black curves are the theoretical predictions based on the zone-folding model described in the text.

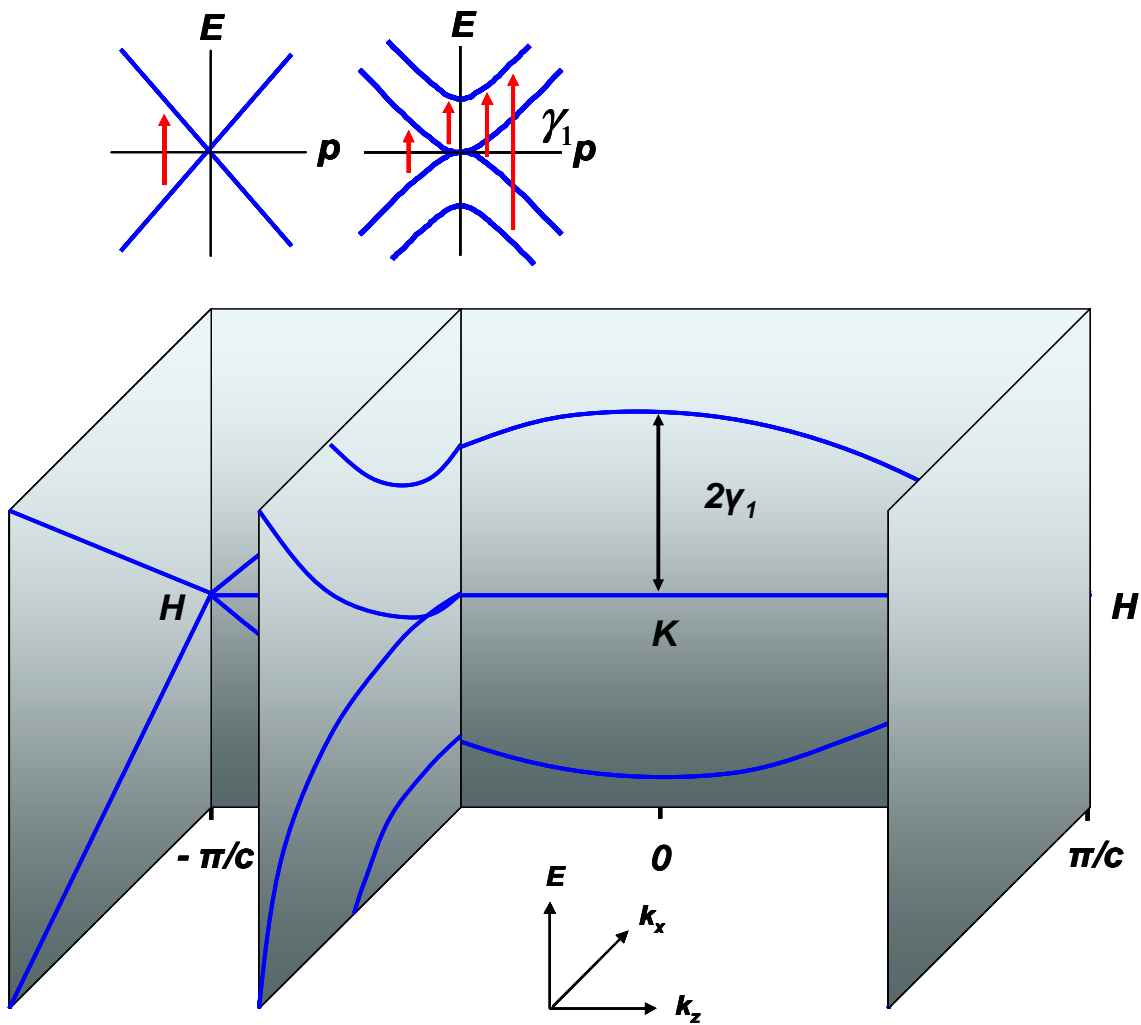

Fig. 4: 3-D low-energy electronic structure of graphite. The cutting planes for single- and bilayer graphene are constructed using the zone-folding scheme presented in the text. The resulting single- and bilayer band structure and the allowed optical transitions are indicated.

CHE-0117752 and ECS-05-07111, the New York State Office of Science, Technology, and Academic Research (NYSTAR); the authors at Brookhaven were supported under Contract DE-AC02-98CH10886 with the U.S. Department of Energy. The synchrotron studies were supported by the National Synchrotron Light Source at Brookhaven and the Center for Synchrotron Biosciences, Case Western Reserve University, under P41-EB-01979 with the National Institute for Biomedical Imaging and Bioengineering.

**Supporting Online Material**
www.sciencemag.org
Supporting online text
Fig. S1

# Supporting Online Material

**The influence of doping on absorption line shape:**

The application of tight-binding theory with the appropriate zone-folding construct allows us to predict the energies of critical points in the band structure of multilayer graphene. These energies match the observed peaks in the optical conductivity spectra for the different graphene multilayers up to $N = 8$ layers [Fig. 3] quite well. The agreement of the predicted line shape of the optical conductivity is, however, less satisfactory. Given the simplicity of the calculation, which neglects, for example, all many-body effects (*1-5*), this is perhaps not surprising. However, as we show here explicitly for the case of the bilayer spectra, we can obtain reasonable agreement within the framework of present theory simply by including a shift in the chemical potential $\mu$, which arises from unintentional doping known to be present in exfoliated graphene samples like ours (*6*).

The discrepancy between the measured bilayer conductivity spectrum and that calculated for the chemical potential at the point of charge neutrality ($\mu = 0$) is clear [Fig. S1]. However, Pauli blocking causes the optical spectra to depend sensitively of the precise value of $\mu$ (*7*). For $\mu = 0$, the contribution to absorption from transitions between the two conduction (and valence) bands is suppressed. This contribution, which gives rise to a strong feature because the pairs of valence and conduction bands are parallel to one another, can be turned on by a shift of $\mu$. A choice of $\mu = 150$ meV yields good agreement between the model and experiment [Fig. S1]. A shift of this magnitude can be expected from spontaneous doping of the graphene layer by interactions with the ambient atmosphere and the substrate (*8*). Similar fits (not shown) were performed for mono- and multilayer graphene and led to improved agreement with experiment. For the case of the bilayer, the predicted sensitivity of the main transition feature to the value of $\mu$ was verified experimentally by studying a field-effect transistor structure [inset of figure and Ref. (*9, 10*)]. A pronounced modification of the optical properties near the principal transitions is induced by changes in the gate potential. Further investigations are needed to determine the role of many-body corrections.

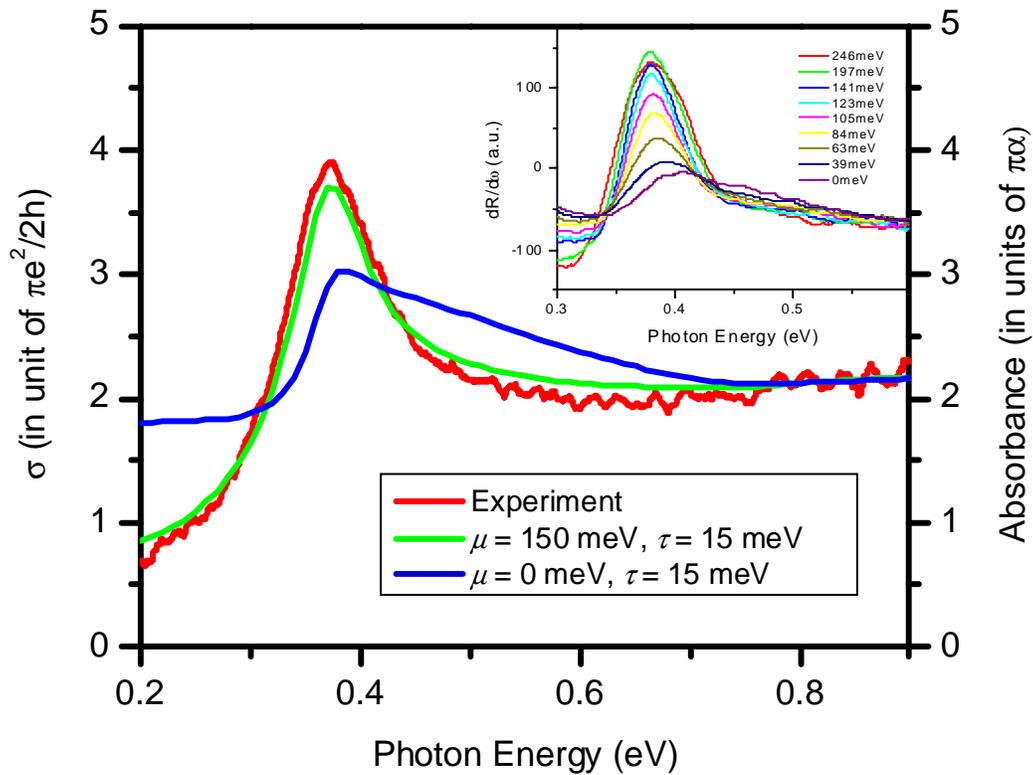

Fig. S1: Comparison of the measured conductivity line shape of bilayer graphene with that calculated from the model presented in the text. The red curve displays the experimental data; blue and green curves are, respectively, theoretical results for samples at room temperature with the chemical potential $\mu$ at and up-shifted by 150 meV from the point of charge neutrality. The high sensitivity of the features of optical conductivity to the chemical potential is demonstrated explicitly in measurements using a bilayer graphene field-effect transistor. The optical response of the bilayer in this structure, presented in the inset as the derivative of the reflectivity spectra with respect to frequency, shows strong variation as we tune the chemical potential of the graphene bilayer by the adjusting gate voltage of the transistor.

**References:**
1. E. G. Mishchenko, *Physical Review Letters* **98**, 216801 (May, 2007).
2. L. Yang, M. L. Cohen, S. G. Louie, *Nano Letters* **7**, 3112 (Oct, 2007).
3. D. Prezzi, D. Varsano, A. Ruini, A. Marini, E. Molinari, *Physical Review B* **77**, 041404 (Jan, 2008).
4. I. F. Herbut, V. Juricic, O. Vafek, *Physical Review Letters* **100**, 046403 (2008).
5. L. Yang, J. Deslippe, C. H. Park, M. L. Cohen, S. G. Louie, *arXiv:0906.0969* (2009).